\documentclass[pra,showpacs,draft,twocolumn,superscriptaddress]{revtex4}

\usepackage[final]{graphicx}
\usepackage{bm}
\begin{document}

\title{Bose-representation for a strongly coupled nonequilibrim
  fermionic superfluid in a time-dependent trap}

\author{I.~V.~Tokatly}
\email{ilya.tokatly@physik.uni-erlangen.de}

\affiliation{Lerhrstuhl f\"ur Theoretische Festk\"orperphysik,
  Universit\"at Erlangen-N\"urnberg, Staudtstrasse 7/B2, 91058
  Erlangen, Germany}

\affiliation{Moscow Institute of Electronic Technology,
 Zelenograd, 124498 Russia}
\date{\today}

\begin{abstract}
Using the functional integral formulation of a nonequilibrium quantum many-body
theory we develop a regular description of a
Fermi system with a strong attractive interaction in the presence of
an external time-dependent potential. In the strong coupling limit
this fermionic system is equivalent to a noequilibrium dilute Bose gas
of diatomic molecules. We also consider a nonequilibrim strongly coupled
Bardeen-Cooper-Schrieffer (BCS) theory and show that
it reduces to the full nonlinear time-dependent
Gross-Pitaevski (GP) equation, which determines an evolution of the
condensate wave function.
\end{abstract}


\maketitle 

A remarkable progress in creation of long-lived cold
diatomic molecules in trapped Fermi gases 
\cite{Regal2003,StrParHul2003,Jochim2003,Cubizolles2003,Durr2004} has
been recently crowned with
experimental demonstration of the molecular Bose-Einstein
condensation (BEC) \cite{Jochim2003a,GreRegJin2003,Zwierlein2003}. One
of the main fundamental goals of these studies is to investigate
experimentally the problem of a crossover from weakly coupled
BCS superfluidity to the molecular BEC. Possibly the
first experimental realization of the crossover regime was 
reported recently in Ref.~\onlinecite{Bartenstein2004}. In the last
decade a theoretical description of the crossover problem also
attracted a considerable interest mainly in the context of hight
temperature superconductivity 
\cite{MelRanEng1993,GTokJETP1993:e,PisStr1996,EngRanMel1997,StiZwe1997,PieStr2000}.
The results of these works show that if the interaction supports
two-particle bound states and the density is low enough,
the system behaves as a dilute Bose gas of diatomic molecules. This
regime corresponds to a strongly coupled fermionic superfluid and a molecular
side of the above crossover. The description of a spatially inhomogeneous
condensed Bose gas is commonly based on GP
equation for the condensate wave function (see
Refs.~\onlinecite{DalGioPit1999,PethickSmith} and references 
therein). Generalization of the previous theoretical works 
to spatially inhomogeneous Fermi systems
demonstrates that in the strong coupling limit the equilibrium
BCS theory reduces the common stationary GP equation
\cite{PieStr2003}.  

One of important features of the experiments with trapped atomic
systems is that 
measurements are frequently performed under nonequilibrium
conditions. The popular time-of-flight technique represents an
example of experimental methods of this type. Therefore it is
desirable to have a consistent kinetic theory of a spatially
inhomogeneous Fermi system with strong pair correlations. Recently
it has been shown that linear response dynamics of strongly
coupled BCS and bosonic GP systems are also equivalent
\cite{PieStr2003b}. However, physically it seems
to be quite likely that the bosonic description of a strongly coupled
nonequilibrium Fermi system should be valid to any order of
nonlinearity provided that external fields are slow functions of time on the
scale of the inverse molecular binding energy. Despite physical
simplicity of this argument a formal nonequilibrium theory of a
superfluid Fermi system in the strong coupling limit is still lacking.  

The present paper is aimed to fill this gap. We formulate a regular
quantum kinetic description of a fermionic system with a strong
attractive interaction in the presence of time-dependent trapping
potential $U({\bf x},t)$. Specifically we consider a two-component
system that is defined by the following Hamiltonian
\begin{eqnarray}\nonumber
 H &=& \int d{\bf x}_{1}
  \Big\{\sum_{j=1}^{2}\Psi^{\dag}_{j}({\bf x}_{1})
  \left[-\frac{\nabla^{2}}{2m} - \mu + U({\bf x},t)\right]
  \Psi_{j}({\bf x}_{1})\\
&-& \int d{\bf x}_{2}
   \Psi^{\dag}_{1}({\bf x}_{1})\Psi^{\dag}_{2}({\bf x}_{2})
   V({\bf x}_{1}-{\bf x}_{2})
   \Psi_{2}({\bf x}_{2})\Psi_{1}({\bf x}_{1})\Big\}
   \label{1}
\end{eqnarray}
where $\Psi^{\dag}_{j}(\Psi_{j})$ is the creation (annihilation)
operator of a Fermi particle of type $j$, 
$V({\bf x})$ is the interparticle interaction potential and $\mu$
is the chemical potential (the Lagrange multiplier, which is
defined by the condition for conservation of the average number of
particles $N$). For simplicity the mass of particles $m$ is assumed to
be independent of $j$.

\begin{figure}
  \includegraphics[width=0.35\textwidth]{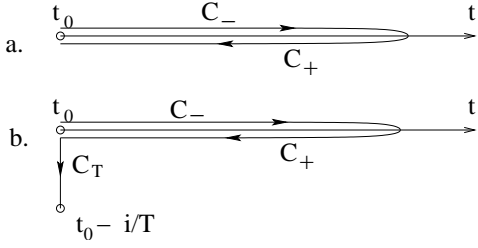}
  \caption{Schematic representation of the integration contours: Contour
    $C'$ which enters general Eq.(\ref{3}) (a); and contour $C$ (b), which
    is used in the generating functional of Eq.~(4)}
  \label{fig1}
\end{figure}

First we briefly summarize some formalities of a nonequilibrium
many-body theory. The complete description of a system with a
time-dependent Hamiltonian $H(t)$ is given by the many-particle density
matrix $\rho(t)$, which satisfies the common equation of motion
$i\partial_{t}\rho(t)=[\rho(t),H(t)]$ with initial condition
$\rho(t_{0})=\rho_{0}$. The average value of any operator $\widehat{O}$
can be calculated as follows
\begin{equation}
\langle \widehat{O}(t)\rangle = \frac{\text{Tr} \rho(t)\widehat{O}}{\text{Tr}
  \rho(t)} =
\frac{\text{Tr}\rho_{0}U^{\dag}(t,t_{0})\widehat{O}U(t_{0},t)}{\text{Tr}
  \rho_{0}},
\label{2}
\end{equation}
where $U(t_{2},t_{1})= T\exp\left\{-i\int_{t_{1}}^{t_{2}}H(t)dt\right\}$
is the evolution operator and $T$ means the usual
chronological ordering. Calculation of the average can be
reformulated in a more compact way if we introduce the following
generating functional
\begin{equation}
Z_{J} = \text{Tr}\rho_{0}T_{C'}\exp\left\{-i\int_{C'}H_{J}(t)dt\right\},
\label{3}
\end{equation}
where $H_{J}(t)=H(t) + J(t)\widehat{O}$ is the Hamiltonian in the
presence of a source field $J(t)$. The time integration in
Eq~(\ref{3}) runs over contour $C'$, which is shown in Fig.~\ref{fig1}a.
Contour $C'$ consists of two branches $C_{-}$ and $C_{+}$ and goes from
initial time $t_{0}$ to infinity ($C_{-}$ branch) and backwards
($C_{+}$ branch). Operator $T_{C'}$ in Eq.~(\ref{3}) orders all times
along the integration contour. It is worth to mention that generating
functional $Z_{J}$,
Eq.~(\ref{3}), is nontrivial (differs from $\text{Tr}\rho_{0}$) only
if the source $J(t^{-})$ at the upper branch differs from the source
$J(t^{+})$ at the lower branch. Using the generating functional of
Eq.~(\ref{3}) one can represent the average of Eq.~(\ref{2}) in terms
of the following functional derivative
$$
\langle \widehat{O}(t)\rangle = i\left[
\frac{\delta \ln Z}{\delta J(t_{-})}\right]_{J=0}.
$$
Introducing sources for any other operator we can calculate the corresponding
observables as well as any higher order correlation function. The
formalism is simplified if the initial condition corresponds
to the thermal equilibrium: $\rho_{0}=exp\{-\beta H(t_{0})\}$,
where $\beta = 1/T$. In this case the generating functional can be
written as a trace of the only chronological exponent
\begin{equation}
Z_{J} = \text{Tr}T_{C}\exp\left\{-i\int_{C}H_{J}(t)dt\right\}
\label{4}
\end{equation}
where all times are ordered along new three-branch contour $C$, which is
shown in Fig.~\ref{fig1}b (for discussion of the three-branch contour
in the context of quantum kinetics see, for example,
Refs.~\onlinecite{RamSmi1986,HorSch1993} and references therein). The main
advantage of Eq.~(\ref{4}) is the possibility to represent it as
a coherent state functional integral. The derivation of this representation
is, in fact, independent of particular time contour. Therefore, it simply
reproduces the corresponding derivation for the partition function in the
equilibrium statistical mechanics \cite{NegeleOrland}. For our
nonequilibrium system we get the result
\begin{eqnarray}
\label{5}
Z_{J} &=& \int_{\psi(t_{0}^{+}-i\beta)=-\psi(t_{0}^{-})}
 \prod_{j=1}^{2}D\psi_{j}^{*}D\psi_{j}
e^{iS[\psi_{j}^{*},\psi]},\\
S &=& \int_{C}dt\int d{\bf x}\sum_{j=1}^{2}\psi_{j}^{*}({\bf x},t)
\widehat{\cal{L}}({\bf x},t)\psi_{j}({\bf x},t) + S_{\text{int}}.
\label{6}
\end{eqnarray}
Here $\psi({\bf x},t)$ is a Grassmann field,
$t_{0}^{-}$ and $t_{0}^{+}-i\beta$ are the initial and the final points of
integration contour $C$ respectively (see
Fig.~\ref{fig1}b). One particle operator
$\widehat{\cal{L}}({\bf x},t)$ in Eq.~(\ref{6}) is defined as follows
\begin{equation}
\widehat{\cal{L}}({\bf x},t) =
i\partial_{t} - \widehat{\cal{H}}({\bf x},t) =
i\partial_{t} + \frac{\nabla^{2}}{2m} - U_{J}({\bf x},t) + \mu.
\label{7}
\end{equation}
For simplicity we introduced only one source $J({\bf x},t)$ for the
density operator and redefined the external potential $U_{J}=U +
J$. The second term, $S_{\text{int}}$, in the action of Eq.~(\ref{6})
is given by the second term in
Eq.~(\ref{1}) with an additional contour integration and with all
$\Psi$-operators being replaced by the corresponding Grassmann fields $\psi$.

Following the standard route we introduce in Eq.~(\ref{5}) Gaussian
integral over nonlocal Bose field $\eta({\bf x}_{1},{\bf x}_{2},t)$
and decouple the four-fermion term $S_{\text{int}}$ in the
action. It is important that field $\eta(t)$ should satisfy boundary
condition $\eta(t_{0}^{+}-i\beta)=\eta(t_{0}^{-})$, which follows from
the boundary condition imposed on Grassmann variables in
Eq.~(\ref{5}). After this Hubbard-Stratonovich transformation
$S_{\text{int}}$ takes the form
\begin{eqnarray}
\nonumber
S_{\text{int}} = &-&\int \big\{
V({\bf x}_{1}-{\bf x}_{2})\eta^{*}({\bf x}_{1},{\bf x}_{2},t)
\psi_{2}({\bf x}_{2},t)\psi_{1}({\bf x}_{1},t) \\
 &+& c.c. + \eta^{*}({\bf x}_{1},{\bf x}_{2},t)V({\bf x}_{1}-{\bf x}_{2})
\eta({\bf x}_{1},{\bf x}_{2},t)\big\}
\label{8}
\end{eqnarray}
with the integration over all internal variables.

Obviously, if we now take the integral over field $\eta$ in the
stationary phase approximation, we end up with a nonequilibrium (and
nonlocal due to general form of the interaction potential)
version of BCS theory. We shall, however, postpone this until the very
end and proceed further at the formally exact level.

Calculating Gaussian integral over Gassmann variables we obtain the
following representation of generating functional $Z_{J}$
\begin{equation}
Z_{J} = Z_{J}^{(0)}\int_{\eta(t_{0}^{+}-i\beta)=\eta(t_{0}^{-})}
 D\eta^{*}D\eta
e^{iS_{\text{eff}}[\eta^{*},\eta]}.
\label{9}
\end{equation}
Factor $Z_{J}^{(0)}$ in Eq.~(\ref{9}) corresponds to the noninteracting
Fermi gas, while the effective
action $S_{\text{eff}}$ is given by the expression, which we
present in the obvious structural form
 \begin{eqnarray}
 \nonumber
 S_{\text{eff}}[\eta^{*},\eta] &=& -i\text{Tr}
 \ln\left(1 - \widehat{G}V\widehat{\eta}\right) - \eta^{*}V\eta \\
  &=& i\sum_{l=1}^{\infty}\frac{1}{2l}
 \text{Tr}\left(\widehat{G}V\widehat{\eta}\right)^{2l} - \eta^{*}V\eta,
 \label{10}
 \end{eqnarray}
where we introduced the notations
\begin{equation}
\widehat{G}(t,t') = \left[
\begin{array}{cc}
G(t,t') &       0          \\
  0     & -\overline{G}(t,t')
\end{array}
\right], \quad \widehat{\eta} = \left[
\begin{array}{cc}
    0       &  \eta(t) \\
\eta^{*}(t)    &    0
\end{array}
\right].
\label{11}
\end{equation}
One particle propagator $G(t,t')$ is defined as the inverse of one particle
operator $\widehat{\cal{L}}$, Eq.~(\ref{7}):
\begin{equation}
\left[i\partial_{t} - \widehat{\cal{H}}({\bf x},t)\right]G(t,t') =
\delta_{C}(t-t'),
\label{12}
\end{equation}
where $\delta_{C}(t-t')$ is the delta-function on contour
$C$. The uniqueness of a solution to Eq.~(\ref{12}) is guaranteed by the
boundary condition
$G(t_{0}^{-},t')=-G(t_{0}^{-}-i\beta,t')$. Introducing one particle
contour evolution operator
$U_{C}(t,t')=
T_{C}\exp\left\{-i\int_{t'}^{t}\widehat{\cal{H}}(\tau)d\tau\right\}$ we
represent the solution to Eq.~(\ref{12}) in the following form
\begin{eqnarray}\nonumber
G(t,t')&=&-iU_{C}(t,t_{0}^{-})\big[(1-\widehat{n}(t_{0}))\theta_{C}(t-t') \\
       &-& \widehat{n}(t_{0})\theta_{C}(t'-t)\big]U_{C}(t_{0}^{-},t),
\label{13}
\end{eqnarray}
where $\theta_{C}(t)$ is the contour step-function and operator
$\widehat{n}(t_{0})= [e^{\beta\widehat{\cal{H}}(t_{0})}+1]^{-1}$
corresponds to the Fermi-function at initial time $t_{0}$.

\begin{figure}
  \includegraphics[width=0.4\textwidth]{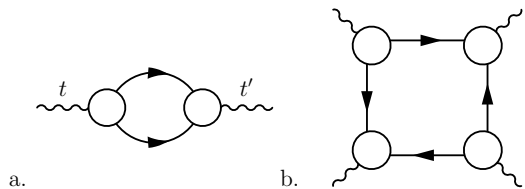}
  \caption{Diagrammatic representation of the second-order (a.) and the
    fours-order (b.) terms in effective action.}
  \label{fig2}
\end{figure}

Let us consider quadratic (Gaussian) part, $S_{\text{eff}}^{(2)}$, of
the effective action, Eq.~(\ref{10}),
\begin{equation}
\label{14}
S_{\text{eff}}^{(2)}=-\int_{C}dtdt'\eta^{*}(t)\{V\delta_{C}(t-t') +
VK_{0}(t,t')V\}\eta(t')
\end{equation}
where
$$
K_{0}({\bf x}_{1},{\bf x}_{2},t;{\bf x'}_{1},{\bf x'}_{2},t') =
iG({\bf x}_{1},t;{\bf x'}_{1},t')G({\bf x}_{2},t;{\bf x'}_{2},t')
$$
is the bare two-particle propagator. Diagrammatic representation of the
second term in Eq.~{\ref{14}} is shown in Fig.~\ref{fig2}a.
Full two-particle propagator $K(t,t')$ in the ladder approximation
satisfies Bete-Salpeter equation
\begin{equation}
\label{15}
K = K_{0} - K_{0}VK.
\end{equation}
Using the definition of Eq.~(\ref{15}) we transform Eq.~(\ref{14}) as follows
\begin{equation}
\label{16}
S_{\text{eff}}^{(2)}=\int_{C}dtdt'\eta^{*}(t)\{K^{-1} -
K^{-1}K_{0}K^{-1}\}\eta(t').
\end{equation}

From this point we concentrate on a strongly coupled superfluid.
We assume that the lowest eigenvalue $-\varepsilon_{0}$ of the
two-particle problem
\begin{equation}
\left[-\frac{\nabla^{2}}{m} - V({\bf r})\right]\chi_{n}({\bf r})=
-\varepsilon_{n}\chi_{n}({\bf r})
\label{17}
\end{equation}
corresponds to a bound state with radius $a_{0}$
($\varepsilon_{0}=1/ma_{0}^{2}$). In the strong coupling limit
$a_{0}$ defines the smallest spatial scale of the problem, which
means that $a_{0}n^{1/3}\ll 1$ and in addition $a_{0}/L\ll 1$,
where $L$ is the characteristic length-scale of external potential
$U({\bf x},t)$. Similarly, the binding energy
$\varepsilon_{0}$ is the largest energy scale, 
which should be large than both temperature $T$ and the inverse
characteristic time related to variations of $U({\bf x},t)$. The
above assumptions allow for significant simplification of the
effective action, Eq.~(\ref{10}). 

In the initial state at $t=t_{0}$ the system is in equilibrium and the
chemical potential is negative and close to half of 
the binding energy $\mu\sim -\varepsilon_{0}/2$ 
\cite{MelRanEng1993,GTokJETP1993:e,EngRanMel1997,StiZwe1997,PieStr2000}.
In fact, the strong coupling limit can be formally viewed as a limit
$\mu\to -\infty$, while the quantity $\varepsilon_{0}+2\mu$ remains
finite. In this limit $n(t_{0})\to 0$, and the one particle
propagator of 
Eq.~(\ref{13}) reduces to the operator of retarded evolution along
contour $C$: $G(t,t')=-iU_{C}(t,t')\theta_{C}(t-t')$. Therefore
the bare two particle Green's function $K_{0}(t,t')$ takes a form of the
two particle retarded propagator
\begin{equation}
K_{0}^{1,2}(t,t') = -i T_{C}e^{-i\int_{t'}^{t}
[\widehat{\cal{H}}_{1}(\tau)+\widehat{\cal{H}}_{2}(\tau)]d\tau}
\theta_{C}(t-t'),
\label{18}
\end{equation}
where indeces 1 and 2 label the propagating particles. Finally,
the full two particle contour propagator $K(t,t')$ of
Eq.~(\ref{15}) can be found as a solution to the following equation 
\begin{widetext}
\begin{equation}
\left[i\partial_{t} + \frac{\nabla_{\bf x}^{2}}{4m} 
- U_{J}\left({\bf x}+\frac{\bf r}{2},t\right) 
- U_{J}\left({\bf x}+\frac{\bf r}{2},t\right)
+ \frac{\nabla_{\bf r}^{2}}{m} + V({\bf r}) + 2\mu+i0\right]K =
\delta({\bf x}-{\bf x'})\delta({\bf r}-{\bf r'})\delta_{C}(t-t').
\label{19}
\end{equation}
\end{widetext}
In Eq.~(\ref{19}) we introduced the notations 
${\bf x}=({\bf x}_{1}+{\bf x}_{2})/2$ and ${\bf r}={\bf x}_{1}-{\bf
  x}_{2}$ for the center-of-mass and the relative coordinates
respectively. The infinitesimally small term, $i0$, in Eq.~(\ref{19}) means
that this equation should be solved with retarded boundary
conditions. By definition, the expression in square brackets in
Eq.~(\ref{19}) is the inverse operator $K^{-1}$, which enters
Gaussian part, Eq.~(\ref{16}), of the effective action.   

To derive a low energy form of the Gaussian action, Eq.~(\ref{16}), we
expand fluctuating Bose field $\eta({\bf x}_{1},{\bf x}_{2},t)$ in
terms of eigen functions $\chi_{n}$ of the two-body problem Eq.~(\ref{17})
\begin{equation}
\eta({\bf x}_{1},{\bf x}_{2},t)= 
\sum_{n}\varphi_{n}({\bf x},t)\chi_{n}({\bf r}).
\label{20}
\end{equation}
Substitution of this expansion into Eq.~(\ref{16}) leads to the result
\begin{equation}
\label{16a}
S_{\text{eff}}^{(2)}=\int_{C}\sum_{m,n}\varphi_{m}^{*}
\langle\chi_{m}|K^{-1} - K^{-1}K_{0}K^{-1}|\chi_{n}\rangle\varphi_{n}.
\end{equation}

Let us calculate the low energy form of matrix elements in Eq.~(\ref{16a}).
The leading contribution to the low energy action is given by the
term $\sim \varphi_{0}^{*}\varphi_{0}$. 
Using the explicit form of $K^{-1}$ (operator in the brackets in
Eq.~(\ref{19})) and Eq.~(\ref{17}) we obtain the following expression
for the matrix element
$\langle\chi_{0}|K^{-1}|\chi_{0}\rangle$ 
\begin{equation}
\langle\chi_{0}|K^{-1}|\chi_{0}\rangle \approx
i\partial_{t} + \frac{\nabla_{\bf x}^{2}}{4m} - 2U_{J}({\bf x},t)
+ 2\mu + \varepsilon_{0}.
\label{21}
\end{equation}
In the derivation of Eq.~(\ref{21}) we have used the fact that
function $\chi_{0}({\bf r})$ is localized on the scale $a_{0}$, which
by basic assumption satisfies the condition $a_{0}/L\ll 1$. To the
same level of accuracy all off-diagonal matrix elements
$\langle\chi_{n\ne 0}|K^{-1}|\chi_{0}\rangle$ vanish due to
orthogonality of functions $\chi_{n}$ with different quantum
numbers. The contributions 
of the type $\langle\chi_{n}|K^{-1}K_{0}K^{-1}|\chi_{0}\rangle$
describe corrections to Eq.~(\ref{21}) (for $n=0$) and the coupling of
the lowest and excited two-particle states (for $n\ne 0$).   In
the low energy corner these
terms are also small since they contain a factor
$K_{0}\sim 1/\mu\sim 1/\varepsilon_{0}$ which is the smallest
parameter of the theory. Matrix elements 
$\langle\chi_{n}|K^{-1}-K^{-1}K_{0}K^{-1}|\chi_{m}\rangle$ with
$n,m\ne 0$ define contributions of excited states to the effective
action. They are at least of the order of the excitation energy
$\varepsilon_{0}$. Since the excited states are decoupled from the
lowest bound state, they do not change low energy physics. Actually
they do contribute only to renormalization of the boson-boson
interaction (see below). It is worth 
noting that in the strong coupling limit the ideal gas contribution
$Z_{J}^{(0)}$ to the generating 
functional of Eq.~(\ref{9}) is also irrelevant as it corresponds to the
ideal Fermi gas with a large negative chemical
potential. Therefore the low energy contribution to the Gaussian part
of the effective action takes a clear physical form
\begin{equation}
S_{\text{eff}}^{(2)}=\int_{C}dtd{\bf x}\varphi^{*}
\left[i\partial_{t} + \frac{\nabla_{\bf x}^{2}}{2M} 
- 2U_{J}({\bf x},t) +\lambda\right]\varphi,
\label{22}
\end{equation}
which corresponds to the Keldysh action of an ideal nonequilibrium
gas of Bose particles with the mass $M=2m$ and the chemical potential 
$\lambda = 2\mu + \varepsilon_{0}$ in the presence of the effective
external potential $U^{B}_{J}({\bf x},t)=2U({\bf
  x},t)$. Equation~(\ref{22}) is, in fact, one of the main formal results of
the present paper.

Terms with $l>1$ in the series in Eq.~(\ref{10}) describe
$l$-boson interactions. Diagrammatic representation for the two-boson
term is shown in Fig.~\ref{fig2}b. Wiggled lines in this figure
correspond to fluctuating Bose fields $\varphi({\bf x},t)$, circles
are boson-fermions ``vertices'' $\Lambda_{0}$, which depend only on relative
coordinate ${\bf r}={\bf x}_{1}-{\bf x}_{2}$ (see Eqs.~(\ref{8}) and
(\ref{20})) 
\begin{equation}
\Lambda_{0}({\bf r})= V({\bf r})\chi_{0}({\bf r})=
\left[-\frac{\nabla^{2}}{m} + \varepsilon_{0}\right]\chi_{0}({\bf r}),
\label{23}
\end{equation}
whereas solid lines stand for the one particle fermionic Green's
functions $G$. Physically interaction of Fig.~\ref{fig2}b corresponds
to a scattering via virtual decay of two composite bosons with subsequent
exchange by the constituent fermions. The main contribution to
internal integrals in the diagram of Fig.~\ref{fig2}b comes from a high
energy region. In fact, the excitation energy $\varepsilon_{0}$ sets an
effective lower cut off for these integrals. Therefore all
effects of inhomogeneity and deviations from the equilibrium, which enter
the diagram via the one particle Green's functions $G$, are
irrelevant for the calculation of this term. The above arguments are
also applicable to all terms with
higher $l$. To the leading order in the strong coupling limit all
coefficients in terms with $l>l$ simply coincide with those obtained in the
equilibrium theory. This should be contrasted to the Gaussian part
($l=1$) where the 
high energy contribution to the diagram of Fig.~\ref{fig2}a is
canceled out by the term $\eta^{*}V\eta$ in the effective action of
Eq.~(\ref{10}). What is left after this cancellation gives exactly
the action, Eq.~(\ref{22}), of nonequilibrium Bose gas in the presence
of the external time dependent inhomogeneity.  

For colliding composite particles of low energy (smaller than
$\varepsilon_{0}$) the diagram Fig.~\ref{fig2}b is independent of
four-momenta. The corresponding constant defines the strength
,$g_{2}^{(0)}$, of an effective short range two-body interaction. Using
Green's functions of the homogeneous 
system as the solid lines in Fig.~\ref{fig2}b and performing frequency
integration we arrive at the result (see, for example, similar calculations in
Refs.~\onlinecite{GTokJETP1993:e,PieStr2000})  
\begin{equation}
g_{2}^{(0)} = 2\sum_{\bf p}
\frac{\Lambda_{0}^{4}({\bf p})}{\left(\frac{{\bf p}^{2}}{m}
+\varepsilon_{0}\right)^{3}}
            = 2\sum_{\bf p}
\left(\frac{{\bf p}^{2}}{m}+\varepsilon_{0}\right)\chi_{0}^{4}({\bf p}).
\label{24}
\end{equation}
In the second equality in Eq.~(\ref{24}) we make an explicit use of
Eq.~(\ref{23}), which relates the boson-fermion
vertex $\Lambda_{0}$ to the bound state wave function
$\chi_{0}$. Similarly one can calculate all higher order bare
coupling constants
\begin{equation}
g_{l}^{(0)} = (-1)^{l}\frac{(2l-1)!}{[(l-1)!]^{2}} \sum_{\bf p}
\left(\frac{{\bf p}^{2}}{m}+\varepsilon_{0}\right)\chi_{0}^{2l}({\bf p}).
\label{25}
\end{equation}
It is, however, well established that in the strong coupling limit the 
pairwise interaction always dominates for any
dimension $d$ of space  \cite{GTokJETP1993:e,PisStr1996}. Consequently
the unrenormalized action for low energy fluctuating Bose fields of
fermionic pairs take the form
\begin{eqnarray}\nonumber
S_{\text{eff}} = \int_{C}dtd{\bf x}\Big\{\varphi^{*}
\Big[i\partial_{t} &+& \frac{\nabla_{\bf x}^{2}}{2M} 
- 2U_{J}({\bf x},t) +\lambda\Big]\varphi\\
&-&  
\frac{1}{2}g_{2}^{(0)}(\varphi^{*}\varphi)^{2}
\Big\}.
\label{26}
\end{eqnarray}

Let us consider 3d system with a short range interaction potential
$V({\bf r})$. If the characteristic range of interaction $R$ is much smaller
than zero energy scattering length $a_{F}$, all
relevant quantities can be expressed in terms of $a_{F}$.
The bound state energy takes a form 
$\varepsilon_{0}=1/ma_{F}^{2}$, which means that
$a_{0}=a_{F}$. Substituting an explicit form of the normalized bound
state wave function
$$      
\chi_{0}({\bf p})= \sqrt{\frac{8\pi}{a_{F}}}
                   \frac{1}{{\bf p}^{2} + a_{F}^{-2}}
$$
into Eq.~(\ref{24}) we arrive at the well known result
\begin{equation}
g_{2}^{(0)} = \frac{4\pi a_{F}}{m}=\frac{4\pi a_{B}^{(0)}}{M},
\label{27}
\end{equation}
where $a_{B}^{(0)}=2a_{F}$ is the bare (unrenormalized) bosonic
scattering length.  

Nonequilibrium BCS theory corresponds to the
stationary phase approximation for generating functional $Z_{J}$ of
Eq.~(\ref{9}). In the strong coupling limit it is sufficient to
take the stationary point of the integral  
\begin{equation}
Z_{J} = \int_{\varphi(t_{0}^{+}-i\beta)=\varphi(t_{0}^{-})}
 D\varphi^{*}D\varphi
e^{iS_{\text{eff}}[\varphi^{*},\varphi]}
\label{28}
\end{equation}
with $S_{\text{eff}}[\varphi^{*},\varphi]$ defined by Eq.~(\ref{26}). 
Stationary point value $\Phi({\bf x},t)$ of Bose field $\varphi({\bf
  x},t)$ satisfies the time-dependent GP equation
\begin{equation}
i\partial_{t}\Phi = \left[-\frac{\nabla_{\bf x}^{2}}{2M} 
+ 2U_{J}({\bf x},t) 
+ \frac{4\pi a_{B}^{(0)}}{M}|\Phi|^{2} - \lambda\right]\Phi
\label{29}
\end{equation}
with initial condition $\Phi({\bf x},t_{0})=\Phi_{0}({\bf x})$, where
$\Phi_{0}({\bf x})$ is the solution to the stationary GP equation 
\begin{equation}
\left[-\frac{\nabla_{\bf x}^{2}}{2M} 
+ 2U_{J}({\bf x},t_{0}) 
+ \frac{4\pi a_{B}^{(0)}}{M}|\Phi_{0}|^{2} - \lambda\right]\Phi_{0} = 0.
\label{30}
\end{equation}
Function $\Phi({\bf x},t)$ plays a role of the condensate wave
function. The relation of $\Phi({\bf x},t)$ to the nonlocal BCS order
parameter follow Eqs.~(\ref{8}) and (\ref{20})
$$
\Delta({\bf r},{\bf x},t)=V({\bf r})\chi_{0}({\bf r})\Phi({\bf x},t).
$$
In the case of short range interatomic interaction BCS theory becomes
local with the local order parameter of the form
$$
\overline{\Delta}({\bf x},t) = \int\Delta({\bf r},{\bf x},t)d{\bf r} = 
\frac{1}{m}\sqrt{\frac{8\pi}{a_{F}}}\Phi({\bf x},t).
$$
Equations (\ref{29}), (\ref{30}) prove the equivalence of strongly
coupled nonequilibrium BCS theory and time-dependent GP theory. As one
could expect on the physical grounds, the
equivalence goes far beyond the linear response regime. 

Stationary phase approximation on the level of generating functional
Eq.~(\ref{9}) 
(BCS theory) neglects high energy ($\sim\varepsilon_{0}$)
fluctuations of Bose fields $\varphi({\bf x},t)$. To construct a
regular description of low energy physics
one has first integrate the high energy degrees of
freedom out. After the integration we obtain the final 
effective kinetic theory. This theory is formally defined by the generating
functional of Eq.~(\ref{28}) and the effective action of Eq.~(\ref{26})
with $g_{2}^{(0)}$ being replaced by the 
physical interaction constant $g_{2}$. As usual, the calculation of
observables with such an effective field theory requires a proper
regularization procedure (see, for example, \cite{Andersen2003}). The
regularization should take into account the fact that coupling
constant $g_{2}$ in the action is already equal to its physical zero
energy value. Numerical value of the physical (renormalized) 
interaction $g_{2}$ is obtained as a zero energy limit of the scattering
matrix ${\cal T}$, which satisfies the following equation
\begin{equation}
{\cal T} = {\cal T}^{(0)} + {\cal T}^{(0)}{\cal G G}{\cal T},
\label{31}
\end{equation}
where ${\cal G}=(K^{-1} - K^{-1}K_{0}K^{-1})^{-1}$ is the full bosonic
propagator (see Eq.~(\ref{16})), and ${\cal T}^{(0)}$ is the full
momentum dependent bare scattering matrix of Fig.~\ref{fig2}b. Since
the relevant energy scale in Eq.~(\ref{31}) is defined by the binding
energy $\varepsilon_{0}$, slow external potential $U_{J}({\bf x},t)$,
which enters Eq.~(\ref{31}) via ${\cal T}^{(0)}$ and ${\cal G}$, is
again irrelevant. In the homogeneous
3d Fermi system with a short range attraction a numerical solution 
to Eq.~(\ref{31}) has been found by Pieri and Strinati
\cite{PieStr2000}. The result 
reported in Ref.~\onlinecite{PieStr2000} ($g_{2}=4\pi a_{B}/M$ with
$a_{B}\approx 0.75a_{F}$) is somewhat different from the result of an
explicit treatment of the four particle problem ($a_{B}\approx
0.6a_{F}$) \cite{PetSalShl2003}. This discrepancy, though not strong,
still requires a clarification, since Eq.~(\ref{31}) should be formally
equivalent to the four particle scattering problem of
Ref.~\onlinecite{PetSalShl2003}.            

In contrast to the 3d case where $a_{B}\sim a_{F}$ and
$g_{2}\sim g_{2}^{(0)}$, in strongly anisotropic traps the renormalization
effects should be more pronounced. For example, in a quasi-2d
trap the physical coupling constant, which enters an effective 2d GP
equation, should take the form \cite{Popov:e,LieSeiYng2001}
$$
g_{2}=\frac{4\pi}{M|\ln\overline{n}a_{B}^{2}|}.
$$
In this case $a_{B}$ can be quite
different from $a_{F}$ since the bosonic scattering length $a_{B}$ is
proportional to the bound state radius $a_{0}$, which strongly depends
on the transverse size of a quasi-2d trap.

In conclusion we developed a regular quantum kinetic theory of a dilute
Fermi gas with strong pair correlations, which is trapped by a
time-dependent external 
potential. The description of this strongly coupled fermionic
superfluid reduces the theory of a dilute nonequilibrium Bose gas
of diatomic molecules. In particular, we proved the equivalence of
nonequilibrium BCS theory in the strong coupling limit and the full
nonlinear time-dependent GP equation. Despite these results may seem
to be quite obvious physically, we believe that our general approach can
be useful in studying more complicated situations related to new
nonequilibrium phenomena in the problem of BCS-BEC crossover. 


\end{document}